\documentclass[aps,preprint]{revtex4}
\usepackage{graphicx}
\usepackage[scanall]{psfrag}
\usepackage{amssymb}

\begin{document}

%
%
%
%
\def\oti{{\otimes}}
\def\lb{ \left[ }
\def\rb{ \right]  }
\def\tilde{\widetilde}
\def\bar{\overline}
\def\hat{\widehat}
\def\*{\star}
\def\[{\left[}
\def\]{\right]}
\def\({\left(}		\def\BL{\Bigr(}
\def\){\right)}		\def\BR{\Bigr)}
	\def\BBL{\lb}
	\def\BBR{\rb}
%
%
\def\zb{{\bar{z} }}
\def\zbar{{\bar{z} }}
\def\frac#1#2{{#1 \over #2}}
\def\inv#1{{1 \over #1}}
\def\half{{1 \over 2}}
\def\d{\partial}
\def\der#1{{\partial \over \partial #1}}
\def\dd#1#2{{\partial #1 \over \partial #2}}
\def\vev#1{\langle #1 \rangle}
\def\ket#1{ | #1 \rangle}
\def\rvac{\hbox{$\vert 0\rangle$}}
\def\lvac{\hbox{$\langle 0 \vert $}}
\def\2pi{\hbox{$2\pi i$}}
\def\e#1{{\rm e}^{^{\textstyle #1}}}
\def\grad#1{\,\nabla\!_{{#1}}\,}
\def\dsl{\raise.15ex\hbox{/}\kern-.57em\partial}
\def\Dsl{\,\raise.15ex\hbox{/}\mkern-.13.5mu D}
%
%
\def\ga{\gamma}		\def\Ga{\Gamma}
\def\be{\beta}
\def\al{\alpha}
\def\ep{\epsilon}
\def\vep{\varepsilon}
\def\la{\lambda}	\def\La{\Lambda}
\def\de{\delta}		\def\De{\Delta}
\def\om{\omega}		\def\Om{\Omega}
\def\sig{\sigma}	\def\Sig{\Sigma}
\def\vphi{\varphi}
%
%
\def\CA{{\cal A}}	\def\CB{{\cal B}}	\def\CC{{\cal C}}
\def\CD{{\cal D}}	\def\CE{{\cal E}}	\def\CF{{\cal F}}
\def\CG{{\cal G}}	\def\CH{{\cal H}}	\def\CI{{\cal J}}
\def\CJ{{\cal J}}	\def\CK{{\cal K}}	\def\CL{{\cal L}}
\def\CM{{\cal M}}	\def\CN{{\cal N}}	\def\CO{{\cal O}}
\def\CP{{\cal P}}	\def\CQ{{\cal Q}}	\def\CR{{\cal R}}
\def\CS{{\cal S}}	\def\CT{{\cal T}}	\def\CU{{\cal U}}
\def\CV{{\cal V}}	\def\CW{{\cal W}}	\def\CX{{\cal X}}
\def\CY{{\cal Y}}	\def\CZ{{\cal Z}}

\def\rvac{\hbox{$\vert 0\rangle$}}
\def\lvac{\hbox{$\langle 0 \vert $}}
\def\comm#1#2{ \BBL\ #1\ ,\ #2 \BBR }
\def\2pi{\hbox{$2\pi i$}}
\def\e#1{{\rm e}^{^{\textstyle #1}}}
\def\grad#1{\,\nabla\!_{{#1}}\,}
\def\dsl{\raise.15ex\hbox{/}\kern-.57em\partial}
\def\Dsl{\,\raise.15ex\hbox{/}\mkern-.13.5mu D}
%
%
%
\font\numbers=cmss12
\font\upright=cmu10 scaled\magstep1
\def\stroke{\vrule height8pt width0.4pt depth-0.1pt}
\def\topfleck{\vrule height8pt width0.5pt depth-5.9pt}
\def\botfleck{\vrule height2pt width0.5pt depth0.1pt}
\def\Zmath{\vcenter{\hbox{\numbers\rlap{\rlap{Z}\kern
0.8pt\topfleck}\kern 2.2pt
                   \rlap Z\kern 6pt\botfleck\kern 1pt}}}
\def\Qmath{\vcenter{\hbox{\upright\rlap{\rlap{Q}\kern
                   3.8pt\stroke}\phantom{Q}}}}
\def\Nmath{\vcenter{\hbox{\upright\rlap{I}\kern 1.7pt N}}}
\def\Cmath{\vcenter{\hbox{\upright\rlap{\rlap{C}\kern
                   3.8pt\stroke}\phantom{C}}}}
\def\Rmath{\vcenter{\hbox{\upright\rlap{I}\kern 1.7pt R}}}
\def\Z{\ifmmode\Zmath\else$\Zmath$\fi}
\def\Q{\ifmmode\Qmath\else$\Qmath$\fi}
\def\N{\ifmmode\Nmath\else$\Nmath$\fi}
\def\C{\ifmmode\Cmath\else$\Cmath$\fi}
\def\R{\ifmmode\Rmath\else$\Rmath$\fi}

\def\barray{\begin{eqnarray}}
\def\earray{\end{eqnarray}}
\def\beq{\begin{equation}}
\def\eeq{\end{equation}}

\def\no{\noindent}

\title{Quantum statistical mechanics of gases
in terms of dynamical filling fractions
and scattering amplitudes}
\author{Andr\'e  LeClair}
\affiliation{Newman Laboratory, Cornell University, Ithaca, NY} 
\date{2006}

\bigskip\bigskip\bigskip\bigskip

\begin{abstract}

We develop a finite temperature field theory formalism in any dimension 
that has the filling fractions as the basic dynamical
variables.  The formalism efficiently decouples zero
temperature dynamics from the quantum statistical sums. 
The zero temperature `data'  is the scattering amplitudes.  
A saddle point condition  leads to an integral
equation which is similar in spirit to the thermodynamic Bethe ansatz
for integrable models, and effectively resums infinite classes of
diagrams.   We present both relativistic and non-relativistic versions.

\end{abstract}


\maketitle

\def\om#1{\omega_{#1}}
\def\Tr{\rm Tr} 
\def\free{\CF} 
\def\xvec{{\bf x}}
\def\kvec{{\bf k}}
\def\kvecp{{\bf k'}}
\def\omk{\om{\kvec}} 
\def\dk{\frac{d^d\kvec}{(2\pi)^d}}
\def\dkind#1{\frac{d^d #1 }{(2\pi)^d}}
\def\dkom{\frac{d^d\kvec}{(2\pi)^d 2 \omk}}
\def\2pid{(2\pi)^d}
\def\fill{{f}}
\def\dkline{\underline{\underline{d\kvec}}}
\def\dklinep{\underline{\underline{d\kvec'}}}
\def\dklineind#1{\underline{\underline{d#1}}}
\def\ket#1{|#1 \rangle}
\def\bra#1{\langle #1 |}
\def\ebH{e^{-\beta H}}
\def\vol{V}
\def\vhat{\hat{v}} 
\def\bfN#1{{\bf{#1}}}
\def\ketbf#1{|{\bf #1}\rangle}
\def\brabf#1{\langle {\bf #1} | }
\def\u{u}
\def\dLR{{{\stackrel{\leftrightarrow}{\d} }}}
\def\Re{{ \it Re}}
\def\Im{{\it Im}}
\def\deltak{\delta^{(\kvec)}}
\def\deltaE{\delta^{(E)}}
\def\n{n}
\def\Fhat{\digamma}
\def\sto{;}
\def\one{{\bf 1}}
\def\Fhatone{\Fhat_1}
\def\CFPI{\CF_{\rm 2PI}}
\def\kernel{{\bf K}}
\def\U{U}
\def\K{\kernel}
\def\dtwo{{\textstyle \frac{d}{2} }}
\def\dplustwo{{ \textstyle \frac{d+2}{2} }}
\def\dplusone{{ \textstyle \frac{d+1}{2}}}


\section{Introduction}

For the investigation  of finite temperature quantum field theory,
the standard and well-developed formalism is based on the euclidean
field theory with time compactified on a circle of circumference 
$\beta = 1/T$ where $T$ is the temperature.    In practice,  this involves the 
sums over Matsubara frequencies in perturbation theory\cite{kapusta,lebellac}. 
Though this formalism is very useful for some problems,  such as the 
finite temperature dependence of the effective potential\cite{jackiw}, for
other properties,  such as the equation of state involving the pressure and
density,  it would be desirable to have a formalism that more clearly preserves the
classical picture of a gas of particles at given density and interacting 
via collisions.   

It helps to realize that at least in principle it is possible to decouple the 
zero temperature dynamics and the quantum statistical sums.  The argument is
simple:  the computation of the partition function $Z = \Tr (e^{-\beta H})$ is
in principle possible from the complete  knowledge of the zero temperature
eigenstates of the hamiltonian $H$.   In practice this is rather difficult
and one resorts to perturbative methods such as the Matsubara method, which
unfortunately entangles the zero temperature dynamics from the quantum statistical
mechanics.     However there does exist a beautiful realization of this kind
of decoupling for integrable quantum field theories in 2 spacetime dimensions
due to Yang and Yang\cite{yangyang}, 
which is refered to as the thermodynamic Bethe ansatz (TBA).  
(For the relativistic version, see \cite{ZamoTBA}.)    In this formalism 
the decoupling is manifest:   The free energy is expressed in terms of 
a pseudo-energy,  in fact it takes the free field form,  and the pseudo-energy
is a solution of an integral equation whose kernel depends on the zero temperature
S-matrix.   In fact,  the S-matrix and the free-particle dispersion relation
are the only properties of the theory that are needed as input.   The TBA is 
a very powerful tool for tracking RG flows and for computing the 
conformal central charge at the fixed points,  and this was one of the main
motivations for this work\footnote{In this $2D$ context, the
central charge  $c$ is the coefficient of the conformal free energy 
$\CF = -c \pi T^2 /6$ and $c$ is the same $c$ as appears in the Virasoro
algebra.}.

Dashen, Ma, and Bernstein derived an  expression for the partition 
function in terms of the S-matrix in\cite{Ma}.  
The derivation is very general, doesn't rely on integrability,  and
is valid in any number of spacetime dimensions\footnote{To our knowledge, 
a complete derivation of the thermodynamic Bethe ansatz equations 
from this formalism is not known.  Some  steps in this direction were taken 
in \cite{Thacker}.}.    The main result in \cite{Ma} is rather formal,
and a considerable amount of additional work  is needed to churn it into a 
useful computational
tool.   Though some steps toward  further developing the formalism were taken in
\cite{Ma} and \cite{Ma2},  the program was never completed and appears to be  eventually 
abandoned
in favor of the Matsubara approach. As should be clear in our work,  some
important aspects of the approach were not well understood in the original works.  
  
In this paper,  we show that,  with a new interpretation,    the main
formula of Dashen et. al.,  can serve as the starting point for a finite
temperature formalism based on the physical occupation number densities
and zero temperature scattering amplitudes.  
First, by using the cluster
expansion for the S-matrix in a  way not explointed   in \cite{Ma}, we
are able to resolve certain difficulties previously encountered 
and render the formalism considerably more appealing.  Secondly,
through a Legendre transformation we are able to formulate the 
quantum statistical mechanics  directly in terms of dynamical filling fractions. 
By ``dynamical'' we refer to the property  that they are determined 
from a variational principle, or saddle point condition, as in the TBA.  

We wish to also point out  the work of 
Lee and Yang on quantum statistical mechanics\cite{LeeYang} which preceeded 
the TBA work of Yang and Yang.  That approach   is not based on the S-matrix
but rather on matrix elements of $e^{-\beta H}$ itself.  Since these matrix
elements depend on the temperature, that approach does not disentangle 
the dynamics from the statistical sums.  
Nevertheless, we  found some of the ideas, 
in particular diagramatic description, very useful.

An outline of the paper is the following.  In the next section
we describe  in a completely general way how one formulates
quantum statistical mechanics of gases in terms of dynamical filling
fractions.  In section III the main result of \cite{Ma} is reviewed
and some potential difficulties pointed out with so-called 
type B terms.     In section IV,  we show how the extensivity
of the free energy follows from the cluster decomposition of the
S-matrix, however in a delicate way that actually provides
some constraints on the interpretation.   We argue that the formalism
is only consistent with the cluster decomposition if one passes
to euclidean space.  This also solves the problem with the type B terms.  
In section V we present a diagrammatic description of the formalism.
It must be stressed that the resulting diagrams have nothing to do 
with finite temperature Matsubara/Feynman diagrams.    In section VI
we show that for the saddle point construction of section II, one needs
only consider the 2-particle irreducible diagrams.     The saddle point
equation is an integral equation that 
 automatically resums infinite numbers of diagrams.  
In section VII we consider contributions that come only from
2-body scattering.   Here also the integral equation sums infinite
numbers of `foam' diagrams.   Though we originally had in mind
applications to relativistic field theory at high temperatures, 
our construction does not assume the underlying theory is 
Lorentz invariant.  In section VIII we develop the non-relativistic
case and solve the integral equation for a constant 2-body scattering
amplitude.

In the concluding section we comment on the potential advantages of
our formalism for certain classes of problems.

\section{Free energy as a dynamical functional of filling fractions}

In the scattering description of quantum statistical
mechanics that we will develop,  a momentum space description
is obviously appropriate.  A very physical momentum space description
uses the occupation number densities as the basic 
dynamical variables.  In this section we generally describe
how this can be done and illustrate it for free particles.  

The free energy density (per volume) $\CF$ is defined as 
\beq
\label{L1}
\CF = - \inv{\beta V} \log Z , ~~~~~~~ Z = \Tr ~ e^{-\beta (H-\mu N)} 
\eeq
where $\beta = 1/T, \mu $ are  the inverse temperature and chemical
potential, $V$ is the d-dimensional spacial volume,  and $H$ and $N$
are the hamiltonian and particle number operator.  Since 
$\log Z$ is an extensive quantity, i.e. proportional to the volume, 
the pressure $p$ of the gas is minus the free energy density
since $p = T \d \log Z/\d V = -\CF$.    For now, 
let us assume there is one species of bosonic ($s=1$) or
fermionic ($s=-1$) particle.   Given $\CF (\mu)$, one can compute
the thermally averaged number density $\n$:
\beq
\label{L2}
n = - \frac{\partial \CF}{\partial \mu } \equiv  
\int \dk ~ f(\kvec)
\eeq 
where $\kvec$ is the d-dimensional momentum vector. ($D=d+1$ is the spacetime dimension.)
  The dimensionless
quantities $f$ are sometimes called the filling fractions.    

One can express $\CF$ as a functional of $f$ (and $\mu$) in a meaningful
way with a Legendre transformation.  Define
\beq
\label{L3}
G \equiv \CF (\mu) + \mu \,  n  
\eeq
Treating $f$ and $\mu$ as independent variables,  then 
using eq. (\ref{L2}) one has that 
$\d_\mu G =0$ which implies it can be expressed only in terms of 
$f$ and satisfies $\delta G/ \delta f = \mu$.   

Inverting the above construction shows that there exists 
a functional $\Fhat (f, \mu)$ 
\beq
\label{L4}
\Fhat (f, \mu) = G(f) - \mu \int \dk ~ f(\kvec) 
\eeq
which satisfies eq. (\ref{L2}) and is a stationary point 
with respect to $f$:
\beq
\label{L5}
\frac{\delta \Fhat}{\delta f} = 0
\eeq
The above stationary condition 
is to be viewed as determining
$f$ as a function of $\mu$.   The physical free energy is
then $\CF = \Fhat$ evaluated at the solution $f$ to the above equation.
We will refer to eq. (\ref{L5}) as the saddle point equation since
it is suggestive of a saddle point approximation to 
a functional integral:
\beq
\label{saddle}
Z = \int D f  ~ e^{-\beta V \Fhat (f)} \approx  e^{-\beta V \CF} 
\eeq

Let us illustrate these definitions for a free theory,  in
a way that will be useful in the sequel.     
In a free theory,  the eigenstates of $H$ are multi-particle
Fock space states $|\kvec_1 , \kvec_2 ....\rangle$. 
Let  $\omega_\kvec$ denote the 
one-particle energy as a function of momentum $\kvec$.  
It is well-known that the trace over the multi-particle 
Fock space gives
\beq
\label{L6}
\CF_0 (\mu) = \frac{s}{\beta} \int \dk ~ 
\log \( 1 - s e^{-\beta (\om\kvec - \mu )} \) 
\eeq
From the definition eq. (\ref{L2}) one finds the filling fractions: 
\beq
\label{L7}
f (\kvec) = \inv{ e^{\beta(\om\kvec - \mu)} -s } \equiv f_0 (\kvec) 
\eeq
In order to find the functional $\Fhat (f,\mu)$ one first computes
$G$ from eq. (\ref{L3}) and eliminates $\mu$ to express it in terms 
of $f$ using eq. (\ref{L7}).  One finds
\beq
\label{L8}
\Fhat_0 (f,\mu) = \int \dk \( 
(\om\kvec - \mu ) f - \inv\beta 
\Bigl[ (f+s) \log (1 + sf) - f \log f \Bigr] \) 
\eeq
One can then easily verify that $\delta \Fhat / \delta f = 0$
has the solution  $f=f_0$ and plugging this back into eq. (\ref{L8}) gives
the correct result eq. (\ref{L6})  for $\CF_0$.  

There is another way to view the above construction which 
involves the entropy.  Write eq. (\ref{L8}) as 
\beq
\label{L9}
\Fhat = \CE - \inv\beta \,  \CS
\eeq
where $\CE$ is the first $(\omega -\mu)f$ term in eq. (\ref{L8}), 
which is the energy density,  
and $\CS$ is the remaining term in brackets.   
One can show by a standard counting argument, which  involves 
the statistics of the particles,  that $\CS$ 
represents the entropy density  of a gas of particles.  
(See for instance \cite{Entropygas}.) 

In the sequel, it will be convenient to trade the chemical
potential variable $\mu$ for the variable $f_0$:  
\beq
\label{L10}
\Fhat_0 (f, f_0) = -\inv{\beta} \int \dk 
\(  s \log (1 + sf)  +  f 
\log \( \frac{1+sf}{f} \frac{f_0}{1+s f_0} \)  \)
\eeq
In section VI we will express the corrections to $\Fhat$ 
for an interacting theory  in 
terms of scattering amplitudes.  

\section{Formal expression for $Z$  in terms of the S-matrix}

The trace that defines $Z$ is computed with respect to
a complete set of orthogonal states $|\alpha \rangle$:
\beq
\label{2.2}
\langle \alpha  ' | \alpha  \rangle = \delta_{\alpha' \alpha}, ~~~~~
\one  = \sum_\alpha  |\alpha  \rangle \langle \alpha  | ~~~~~ 
\Longrightarrow  
Z = \sum_{\alpha } \langle \alpha  | ~ e^{-\beta ( H - \mu N )} | \alpha
 \rangle 
\eeq
Let us separate $H$ into  free ($H_0$) and interacting ($H_1$) parts:
\beq
\label{FS1}
H = H_0 + H_1 
\eeq
Since the states $\alpha$ in eq. (\ref{2.2}) are not required to
be eigenstates of $H$, let us take the trace over eigenstates of $H_0$:
\beq
\label{FS2}
H_0 |\alpha \rangle = E_\alpha |\alpha \rangle
\eeq
In the next section we will specialize to plane-wave scattering states,
but for the remainder of this section one need not be so specific.

It was shown how to express the thermal trace in terms of the S-matrix in 
\cite{Ma}.   The necessary algebraic tools are familiar from the formal theory
of scattering\cite{Gottfried,Weinberg}. 
There is a simple derivation of what will turn out to be
the essential  term which goes as follows. 
For simplicity we first set the chemical potential
$\mu$ to zero;  it can easily be restored at the end by letting
$\omega_\kvec \to \omega_\kvec - \mu$.   Define the resolvent operator
\beq
\label{R1}
G(E) = \inv{E - H + i\vep} 
\eeq
where $E$ is a real variable and $\vep$ is small and positive.  If $H$ is hermitian,
then in the limit $\vep \to 0^+$: 
\beq
\label{R2}
G(E) - G(E)^\dagger = - 2\pi i ~ \delta (E-H)
\eeq
Assuming that the spectrum of $H$ has $E\geq 0$, 
one then evidently has
\beq
\label{R3}
Z = -\inv{2 \pi i} \int_0^\infty dE ~ e^{-\beta E} ~ {\rm Tr} \( G(E) - G^\dagger (E) \)
\eeq
Henceforth, we will not always display that operators 
depend on the variable $E$;  all operators depend on $E$ 
except for $H, H_0, H_1$.  

In order to obtain an expression that is meaningful when the trace is taken 
over the free Fock space, one wants to separate out expressions depending on $H_0$
as much as possible.  
The resolvent satisfies  the equation\footnote{$(A-B)^{-1} = A^{-1} ( 1+B (A-B)^{-1}) 
=(1+(A-B)^{-1}B )A^{-1}$.}
\beq
\label{R4}
G = G_0  + G_0  ~ H_1 ~ G 
\eeq
where $G_0 (E) = (E-H_0 + i \vep )^{-1}$.    As in standard discussions of scattering
(see e.g. \cite{Gottfried}),   define the operator 
\beq
\label{R5}
T(E) = H_1 + H_1 ~G(E) ~H_1 
\eeq
It  satisfies
\beq
\label{R6}
H_1 G = T G_0  , ~~~~~ G H_1 = G_0  T 
\eeq
Using eq. (\ref{R6}) one now has
\beq
\label{R7}
G = G_0 + G_0 T G_0 
\eeq
and 
\beq
\label{R8}
Z= Z_0 - \inv{ \pi } \Im  \int_0^\infty ~ dE ~ e^{-\beta E} ~  
{\rm Tr} \( G_0 T G_0  \) 
\eeq
where $Z_0 = {\rm Tr} e^{-\beta H_0}$ is the partition function of the free
theory.  
Using now the cyclicity of the trace, $\d_E G_0 = - G_0^2 $,  an integration by
parts,  and $\Im G_0 = -\pi \delta (E-H_0)$,   one finds:
\beq
\label{decomp}
Z = Z_0 + Z_A + Z_B
\eeq
where 
\beq
\label{ZA}
Z_A \equiv  -\beta \Re\(  \int dE e^{-\beta E} ~ \Tr \(\delta (E-H_0) T(E) \) \)  =
- \beta \sum_\alpha e^{-\beta E_\alpha} ~ \Re ( T_{\alpha;\alpha} )
\eeq

The term $Z_B$  involves $\d_E T = - T G_0^2 T$ and is thus quadratic in $T$.  
The derivation in \cite{Ma} performs further algebraic manipulations of eq.(\ref{R8})
and elegantly expresses the final result for both $Z_{A,B}$  in terms of the S-matrix. 
The construction is summarized in Appendix A, where one finds: 
\beq
\label{S.1}
Z = Z_0 + \inv{4 \pi i} \int_0^\infty dE ~ e^{-\beta E} ~ 
\Tr \( S^{-1} \dLR_E S \) 
\eeq
where  $X \dLR_E Y \equiv X (\d_E Y) - (\d_E X) Y$,
and $S(E)$ is an operator valued function of $E$ related to the S-matrix\footnote{Another
form of eq. (\ref{S.1}) is based on the identity 
$\Tr S^{-1} \dLR_E S = 2 i \Im \Tr \d_E \log S$, where we have used the  on-shell relation
$S^{-1} = S^\dagger$.  Though this form may be useful to compare with the TBA, 
we do not use it in this paper.}.   
More specifically, 
\barray
S(E) &=&  1 - 2 \pi i \, \delta (E- H_0) ~ T(E)
\nonumber 
\\
S^{-1} (E) &=& 1 + 2 \pi i \, \delta (E- H_0) ~ T^\dagger (E) 
\label{S.2}
\earray
The on-shell matrix elements of $S(E)$ are the usual S-matrix elements:
\beq
\label{Son}
\langle \alpha ' | S(E) | \alpha \rangle = \langle \alpha' | \alpha \rangle 
- 2 \pi i \, \deltaE_{\alpha';\alpha} ~ T_{\alpha' \sto  \alpha} 
, ~~~~~~{\rm iff~~}  E = E_\alpha
\eeq
where $\deltaE_{\alpha';\alpha} \equiv \delta (E_{\alpha'} - E_\alpha )$. 
The condition ``${\rm iff~~} E = E_\alpha$'' is what is referred to as ``on-shell''.

A significant amount of work remains in order to obtain a useful calculational tool from
eq. (\ref{S.1}),  and this  will be the main subject of the next section.  
For now,  note  that 
there are two types of terms (in addition to $Z_0$) in eq. (\ref{S.1}) as
in eq. (\ref{decomp}), 
where 
\barray
\label{R18}
Z_A &=& -\inv{2} \int_0^\infty dE ~ e^{-\beta E} ~ \Tr \bigl[  
\d_E \( \delta(E-H_0) (T + T^\dagger ) \) \bigr] 
\\  \nonumber
Z_B &=& -i \pi  \int_0^\infty dE ~ e^{-\beta E}  ~\Tr 
\[ \( \delta(E-H_0) T^\dagger\)  \dLR_E \( \delta(E-H_0) T \)  \] 
\earray
We will refer to these as Type A and Type B terms. 
Because of the $\delta(E-H_0)$ factors,  we can now trace over
eigenstates $|\alpha\rangle $  of $H_0$ and perform the integral over $E$.  
Consider first $Z_A$.     Integrating by parts gives
\beq
\label{R19}
Z_A = - \beta \sum_\alpha  e^{-\beta E_\alpha} ~ \Re \( T_{\alpha;\alpha} \) + 
\deltaE (0) \Re\(  \langle 0 | T | 0 \rangle \) 
\eeq
where $|0\rangle$ is the free-particle vacuum of zero energy.   Assuming
the vacuum is stable,  $\langle 0| T |0 \rangle = 0$ and the 
$\deltaE (0)$ term can be dropped and eq. (\ref{R19}) agrees with
(\ref{ZA}).   ($\deltaE (0)$ will be regularized below.)     

The type A and B terms are quite different.  The type B terms are actually rather
peculiar since they may potentially spoil the extensivity of the free energy
since  they don't obviously have the same connectivity properties as the type A terms. 
We will return to this issue in the next section and actually propose that they
should be discarded.     
$Z_B$ can be simplified using the optical theorem, which follows from  $S^{-1}  S = 1$:
\beq
\label{optical}
T - T^\dagger = - 2 \pi i ~ T^\dagger ~ \delta (E-H_0) ~ T 
\eeq
Using this,  and the cyclicity of the trace, one finds that the terms
involving the derivative of the $\delta$-functions  vanish.  
Inserting two complete sets of states one then finds:
\barray
\label{R20}
Z_B &=& -i\pi \sum_{\alpha, \alpha'} e^{-\beta E_\alpha} ~ \delta(E_\alpha - E_{\alpha'})  
~ T^*_{\alpha'\sto \alpha}   \dLR_{_{E_\alpha}} T_{\alpha'\sto \alpha} 
\\ \nonumber
&=& 2\pi  \sum_{\alpha, \alpha'} e^{-\beta E_\alpha} ~ \delta(E_\alpha - E_{\alpha'})  
~ \Re (T_{\alpha'\sto \alpha})    \dLR_{_{E_\alpha}} \Im (T_{\alpha' \sto \alpha} )
\earray
Note that the Type B terms vanish if $\Im T = 0$.

\def\xvec{{\bf x}}

\section{Extensivity of the free energy and the cluster decomposition}

In this section we specialize  to a trace over
plane wave scattering states and describe some new features that arise. 
In particular, 
in infinite volume,  the free energy is expected to be extensive, i.e. 
proportional to the spacial 
volume $V$.  So, one first must understand how all the various types of terms 
sum up in a way that can be reorganized as the exponential of something proportional
to the volume.   Clearly this has to do with properties of connectedness.  
In the present formalism, this property is essentially  a consequence of the
cluster decomposition  of the S-matrix.  
The volume factors will arise from the following regularization of 
the momentum space delta-function: 
\beq
\label{deltareg}
(2\pi)^d \delta^{(d)} (0) = \lim_{\kvec = \kvec'} \int d^d \xvec
 ~ e^{i \xvec  \cdot (\kvec
-\kvec')} \equiv  V
\eeq
If  the expression (\ref{decomp}) clusters in the expected way,  
then the free energy $-T \log Z$  can be identified with the sum of all terms with only
one power of $V$.   As we will show, {\it requiring} that the cluster decomposition for 
(\ref{decomp})  leads to  a free energy with this property actually provides some
constraints on the interpretation of various terms.

\subsection{Fock space and S-matrix conventions}

Since we are considering a quantum field theory, 
the Hilbert space of the free theory is 
a Fock space.  Let us now fix our normalizations for the
free particle states and their scattering amplitudes. 
The  creation-annihilation operators satisfy:
\beq
\label{2.3}
a_\kvec a^\dagger_\kvecp ~ -s ~ a^\dagger_\kvecp a_\kvec = \2pid \, 
 ~ \delta^{(d)}  (\kvec - \kvecp )
\eeq  
The Hilbert space is then spanned by the multi-particle states
\beq
\label{FS4} 
|\kvec_1  \kvec_2  \cdots  \kvec_N \rangle 
= \( \prod_i \sqrt{2\om{\kvec_i}} \)  a^\dagger_{\kvec_1} 
\cdots  a^\dagger_{\kvec_N} |0 \rangle 
\eeq
satisfying
\beq
\label{FS5}
H_0 |\kvec_1  \kvec_2  \cdots  \kvec_N \rangle 
= \( \sum_i \om{\kvec_i} \)  |\kvec_1  \kvec_2  \cdots  \kvec_N \rangle 
\eeq
(The factors $\sqrt{2 \om{\kvec}}$ are a matter of convention.)
One has the non-zero inner products:
\beq
\label{2.4} 
\langle \kvec'_1 \cdots \kvec'_N | \kvec_1 \cdots \kvec_N \rangle 
= (2\pi)^{Nd} \sum_{\CP} ~ s^p ~ \prod_{i=1}^N 2 \om{\kvec_i} 
\delta(\kvec'_i - \kvec_i ) 
\eeq
where the sum is over the $N!$ permutations $\CP$ of the order of
the $\kvec'_i$ and $p(\CP)$ is the degree of the permutation 
such that $p=0$ ($p=1$) if $\CP$ involves an even (odd) number of
pairwise
permutations of particles.  
The above implies the following resolution of the identity:
\beq
\label{2.5}
\one  = \sum_{N=0}^\infty \inv{N!} \int  
\dklineind{\kvec_1 } \, \dklineind{\kvec_2} \cdots \dklineind{\kvec_N} ~~ 
| \kvec_1 \cdots \kvec_N \rangle \langle \kvec_1 \cdots \kvec_N | 
\eeq
where for convenience we have defined the notation:
\beq
\label{2.6}
\int \dkline \equiv \int  \frac{d^d \kvec}{\2pid  } \inv{2 \om{\kvec}}
\eeq
(We have chosen our normalization of states so 
that the above integration measure over $\kvec$  is Lorentz 
invariant.  We emphasize however that we are not assuming the theory to be
Lorentz invariant;  a non-relativistic case is worked out in section VIII.) 

\def\M{\CM}
\def\teehat{{\hat{T}}}

\subsection{Cluster decomposition}

In order to simplify the notation, 
the free particle states $\ket{\kvec_1 , \kvec_2 , ..., \kvec_n}$
will be denoted as $\ketbf{12...n}$ and the S-matrix elements as
\beq
\label{C1}
\bra{\kvec'_1, \kvec'_2, ..., \kvec'_n } S  \ket{\kvec_1, \kvec_2, ..., \kvec_m}
= 
S_{\bf 1'2'\cdots n' \sto  12\cdots m} 
\eeq
For the S-matrix the  cluster decomposition 
is based on the physical requirement that particles that are
causally separated cannot scatter (see for instance \cite{Weinberg}). 
The cluster decomposition also ensures that the free energy only
depends on connected S-matrix elements.   
The cluster decomposition may be expressed as follows:
\beq
\label{C2}
\bra{\alpha'} S \ket{\alpha} = 
\sum_{\rm partitions} s^p S^c_{\alpha_1' \sto \alpha_1} \, 
S^c_{\alpha_2' \sto \alpha_2 } \, S^c_{\alpha_3 ' \sto \alpha_3} \cdots
\eeq
where the sum is over partitions of the state $|\alpha  \rangle$ 
into clusters $|\alpha_1\rangle, |\alpha_2 \rangle, ....$.
(The number of particles in $|\alpha_i\rangle$ and 
$\langle \alpha'_i |$ is not necessarily the same.)    
The above formula essentially {\it defines} what is
meant by the connected matrix elements $S^c$.  
The particles are assumed to be stable which implies
\beq
\label{C3}
\bra{\kvec'_1 } S \ket{\kvec_1}  = \bra{\kvec'_1} {\kvec_{1} } \rangle 
=  
(2\pi)^d 2 \omega_{\kvec_1} \delta^{(d)} (\kvec'_1 - \kvec_1) 
\equiv \deltak_{\bf 1' 1} 
\eeq
It will be convenient  to express the cluster decomposition in
terms of $\teehat$ defined by:
\beq
\label{teehateq}
S = 1 + i \teehat
\eeq
The cluster decomposition for $\teehat$ is then  same as for that of $S$
but without the terms involving only delta functions which come from the
`$1$' in $S= 1 + i \teehat$.  (One can show that these additional terms 
involving only delta functions are what give rise to $Z_0$.) 
The connected matrix elements $\teehat^c$ are characterized as
having only a single overall momentum and energy conserving delta function
and cannot be factorized into functions of only a subset of the momenta. 
Below  we will  need  $\teehat^c$ in terms of the conventional scattering amplitudes 
$\CM$:
\beq
\label{Type1.1} 
\teehat^c_{\alpha'\sto \alpha} = - 2\pi \deltaE_{\alpha'; \alpha} T_{\alpha';\alpha}
=  (2\pi)^{d+1} \deltaE_{\alpha' ; \alpha  } \delta^{(d)} 
(\kvec_\alpha - \kvec_{\alpha'} ) ~ \M_{\alpha' \sto \alpha} 
\eeq
where $E_\alpha, \kvec_\alpha$ are the total energy and momentum of the state $|\alpha\rangle$
and $\deltaE_{\alpha' ; \alpha} \equiv \deltaE (E_{\alpha'} - E_\alpha )$. 

For $2\to 2$ and $3\to 3$ particles the cluster decomposition 
for $\teehat$  then reads: 
\barray
\nonumber
\teehat_{\bf 1' 2' \sto 1 2} &=& \teehat^c_{\bf 1' 2' \sto 12}  
\\ 
\label{C4} 
\teehat_{\bf 1'2'3' \sto 123} &=& \teehat^c_{\bf 1'2'3'\sto 123} 
+ \[ \deltak_{\bf 1'1} ~\teehat^c_{\bf 2'3'\sto23} 
\pm {\rm perm.} \]_9 
\earray
The subscript $_9$ indicates the number of permutations within the brace. 
If the one-particle states were not stable, there would be additional terms.  
In order to illustrate some important additional features,
we will also need the $4 \to 4$ particle decomposition:
\barray
\nonumber
\teehat_{\bf 1'2'3'4'\sto1234} &=&  \teehat^c_{\bf 1'2'3'4'\sto1234} 
+  i \[\teehat^c_{\bf 1'2'\sto12} \teehat^c_{\bf 3'4'\sto34} \pm {\rm perm.} \]_{18} 
+ \[ \deltak_{\bf 1'1} ~ \teehat^c_{\bf 2'3'4'\sto234} \pm {\rm perm.} \]_{16}
\\ 
\label{C5}
&~& ~~~~~~~~~~~~~~~~~~~+ 
\[\deltak_{\bf 1'1} \deltak_{\bf 2'2} ~ \teehat^c_{\bf 3'4'\sto34} \pm {\rm perm.} \]_{72}
\earray

\def\perm{{\rm perm.}}

\subsection{One-particle resummation and emergence of filling fractions}

To compute the required trace,  one must set $\{ \kvec' = \kvec\}$.  This
leads to a more specialized cluster decomposition that is suitable to 
compute $Z$.   For instance,  the 9 terms in braces in eq. (\ref{C4}) separate into
$3+6$ distinct types of terms depending on whether they contain $\delta^{(d)} (0)$: 
\beq
\teehat_{\bf 123; 123} = \teehat^c_{\bf 123,123} + 
\[ (2\pi)^d \delta^{(d)} (0) 2\om{\kvec_1}
 ~ \teehat^c_{\bf 23;23} \pm {\rm perm.}\]_3 
+ \[ s \deltak_{\bf 12} \teehat^c_{\bf 23;13} \pm \perm \]_6
\eeq

In order to illustrate the variety of terms that can arise from the cluster
decomposition,  let us compute the Type A   terms for low numbers of
particles. For simplicity we set the chemical potential to zero.   
For two particles one finds:
\beq
\label{Low.1}
Z_A ~\Bigr|_{\rm 2~ part.} = \frac{\beta V}{2} \int \dklineind{\kvec_1} \, 
 \dklineind{\kvec_2} 
~ e^{-\beta (\omega_1 + \omega_2 )} Re (\CM_{\bf 12;12} )
\eeq
where we have used eq. (\ref{deltareg}).  ($\omega_i \equiv \omega_{\kvec_i}$)

For 3 particles one finds
\barray
\label{Low.2.a}
Z_A ~\Bigr|_{\rm 3~ part.} &=&
\frac{\beta V}{3!} \int \dklineind{\kvec_1} \, \dklineind{\kvec_2} \, \dklineind{\kvec_3} 
~ e^{-\beta(\omega_1 + \omega_2 + \omega_3)} ~ \Re (\CM_{\bf 123;123}) 
\\
\label{Low.2.b}
&~& ~~~~~ + s \frac{\beta V}{2} \int  \dklineind{\kvec_1} \,  \dklineind{\kvec_2} ~
e^{-\beta(\omega_1 + \omega_2 )} (e^{-\beta \omega_1} + e^{-\beta\omega_2} ) \Re (\CM_{\bf 12;12})
\\
\label{Low.2.c}
&~& ~~~~~~~~~~ + \beta V^2 
\( \int \dklineind{\kvec_1}\,  \dklineind{\kvec_2} ~ e^{-\beta(\omega_1 + \omega_2 )} ~ 
\Re (\CM_{\bf 12;12}) \)
\( \int \frac{d^d \kvec}{(2\pi)^d}  ~ e^{-\beta \omega_{\kvec}} \) 
\earray
These terms have the following interpretation.  Eq. (\ref{Low.2.a}) is a new contribution of
the same kind as eq. (\ref{Low.1}).   The term (\ref{Low.2.b}) just modifies the 
integration measure $\dklineind{\kvec}$ for the (\ref{Low.1}) term.  The combined measure
factors are the first terms in the expansion of the free filling fraction $f_0$ 
defined in eq. (\ref{L7}).  
This manner in which the filling fraction emerges
was also a feature of one approach to finite temperature
correlation functions in \cite{LecSal,LecMuss}. 
 The last term is proportional to $V^2$ and is 
thus expected to be a $V^2$ term in the expression $Z= \exp (-\beta V \CF)$.  One can easily
verify that it is the correct combinatorial 
product of a term from the free contribution $\CF_0$ and another 
from the two-particle contribution in eq. (\ref{Low.1}).

A simple combinatorial argument shows that one can sum up all the terms that are a product   of 
one $\teehat^c$ and some  $\deltak_{\bf nm}$ factors to obtain 
\beq
\label{Low.3}
\CF = \CF_0 -  \sum_{N=2}^\infty \inv{N!} \int 
\( \prod_{n=1}^N \dkind{\kvec_n} \frac{f_0 (\kvec_n)}{2\om{\kvec_n}}  \) ~ 
\Re \( \CM_{\bf 12..N; 12..N} \)  + .....
\eeq
Restoring the chemical potential $\mu$,  the dependence on it is only
through $f_0$ as given in eq. (\ref{L7}).  

For a scalar field interacting with potential 
$V(\phi ) = \lambda \phi^4 / 4!$, to lowest order $\CM_{\bf 12;12} = -\lambda$\cite{Peskin}
and the 2-particle contribution in eq. (\ref{Low.3}) gives the same result to 
order $\lambda$, as the Matsubara approach\cite{kapusta}.  
(In the latter approach
this arises as a two-loop finite temperature Feynman diagram.)
The $3\to 3$ particle contribution to eq. (\ref{Low.3}) 
was also shown to agree with the 3-loop result in \cite{Bugrij}.

\subsection{The need to continue to euclidean space}

There are additional  contributions to $\CF$  that come from terms in
the cluster expansion that involve more than one $\teehat^c$ factor.  They first arise 
at four particles.
The terms in $\teehat_{\bf 1234;1234}$ that give new contributions to $\CF$ not
already included in eq. (\ref{Low.3}) are the following:
\beq
\label{Low.4}
\teehat_{\bf 1234;1234} = 
i\[ \teehat^c_{\bf 12;12} \teehat^c_{\bf 34;34} \]_{3} 
+i \[ s \teehat^c_{\bf 13;12} \teehat^c_{\bf 24;34} \]_{12}
+ i \[ \teehat^c_{\bf 12;34} \teehat^c_{\bf 34;12} \]_3 + ....
\eeq
In order to compute their contribution to $Z_A$, we start with the expression
(\ref{ZA}) in terms of $T$.  We then obtain the cluster decomposition of
$T$ from eq. (\ref{Low.4}) using 
$\teehat_{\alpha' ; \alpha} = - 2\pi \deltaE_{\alpha';\alpha} ~ T_{\alpha';\alpha}$.
In carrying this out, 
 one sees that one can  divide both sides of eq. (\ref{Low.4}) by $\deltaE (0)$
and obtain something well-defined.    Finally we express the final result in terms of
the scattering amplitudes $\CM$.  
The first term in eq. (\ref{Low.4}) in this way leads to: 
\beq
\label{delta0}
Z_A ~\Bigr|_{\rm 4 ~ part.} = ..... \frac{\beta V^2}{8} \Re \[ 2 \pi i \deltaE (0) 
\( \int \dklineind{\kvec_1}\, \dklineind{\kvec_2} e^{-\beta (\om1 + \om2)} ~ 
\CM_{\bf 12;12} \)^2 
\]  +.....
\eeq
Since this is proportional to $V^2$ it must come from the square of 
the 2-particle term in the expansion  of  
$Z = (1 + ...) \exp( \beta V/2  \int \Re (\CM_{\bf 12;12}) )$. 
In attempting to match this with eq. (\ref{delta0}), one notices 
a very interesting phenomenon:  the two terms can only be identified 
 after one makes the Wick rotation
\beq
\label{wick}
\deltaE_{\alpha' ; \alpha} \to -i \deltaE_{\alpha' ; \alpha}
\eeq  
One then regularizes the $\deltaE (0)$ as follows:
\beq
\label{deltaEreg}
2\pi \deltaE (0) \equiv  \beta 
\eeq
The above equation can be understood as  following from the fact that
in the Matsubara approach to finite temperature field theory,  one
goes to euclidean space and compactifies the time on a circle of 
circumference, i.e. volume,  $\beta$.  The above equation is then simply the
euclidean time version of eq.   (\ref{deltareg}).

Implementing the rules discussed in the last paragraph,  the two additional 
sets of terms in the cluster decomposition (\ref{Low.4}) lead to two new
contributions to the free energy.   In the resummation of  1-particle
terms,  the filling fractions $f_0$ again emerge. 
These lead to:
\barray
\label{Low.5.a}
\CF &=& - \frac{s \beta}{2} 
 \int \dkind{\kvec_1}\,  \dkind{\kvec_2}\, \dkind{\kvec_3} ~
\frac{f_0 (\kvec_1)}{2\om{1}} \frac{ f_0 (\kvec_2)^2}{4\om{2}^2} 
\frac{ f_0 (\kvec_3)}{2\om3} 
 \Re ( \CM_{\bf 12;12} \CM_{\bf 23;23} )
\\
\label{Low.5.b}
&~&~-\inv{8} \int \dkind{\kvec_1}\,  \dkind{\kvec_2} \, \dkind{\kvec_3}
\( \prod_{n=1}^4 \frac{f_0 (\kvec_n )}{2\om{n}}  \)
2 \pi  \deltaE_{12;34}  
\Re ( \CM_{\bf 12;34} \CM_{\bf 34;12} ) +...
\earray
where $\kvec_4 = \kvec_1 + \kvec_2 - \kvec_3$ and $\om4 = \om{\kvec_4}$.

\subsection{$Z_B$ terms:  To B  or not to B?}

We now return to issue of the type B terms in eq. (\ref{R20}).  
It is clear from the previous results of this section that
the property of the extensivity of the free energy arises
in a delicate way from the cluster decomposition.  Since
$Z=Z_0 + Z_A + Z_B$ and 
the $Z_B$ terms are quadratic in $T$ they could very easily
spoil the extensivity.  We have already shown how 
$Z_0 + Z_A$ exponentiates to something proportional to the volume,
but the way this comes about is quite subtle. 

We now argue that the need to go to euclidean space cures this
potential problem in a very satisfying way.   In trying to perform
the match discussed in the last subsection,  one actually needs
a stronger constraint:  
\beq
\label{stronger}
\beta \( \int \Re (\CM_{\bf 12;12}) \)^2 = 
\Re \[  2\pi \deltaE (0) \( \int \CM_{\bf 12;12} \)^2 \] 
\eeq
I.e. $\CM_{\bf 12 ; 12}$ must be real.   It is well-known that
the imaginary parts of scattering amplitudes arise through cuts
when one is in Minkowski space\cite{Peskin}.  In euclidean space
the imaginary part vanishes.  Therefore,  the $Z_B$ terms are zero.

\section{Diagrammatic description}

The contributions to $\CF$ have a nice  diagrammatic
description.  Let us represent the scattering 
amplitudes $\CM_{\bf 12..m; 12..n}$
as a vertex with $n$ incoming and $m$ outgoing lines.  
We construct a diagram with no external lines by linking the lines of vertices.    
The rules for computing a contribution to the free energy density $\CF$ are then
the following:

\no
(i)   Assign a factor of the scattering amplitude and 
an energy conserving delta function  $2\pi \deltaE_{\alpha';\alpha} \CM_{\alpha' ;\alpha}$ 
to each
vertex as in Figure 1. 

\no (ii) Assign a factor $f_0 (\kvec) / 2 \om\kvec$ to each line as in Figure 1.   

\no(iii) Conserve momentum $\kvec$  at each vertex and integrate over  every unconstrained 
momentum with 
$\int d^{(d)} \kvec / (2\pi )^d $. 

\no (iv)  Identify $2\pi \deltaE (0) = \beta$.

\no (v)  Divide by the symmetry factor of the diagram, defined as the number of
permutations of the internal lines that do not change the topology of the graph,
including relative positions.  

\no  (vi) For fermions determine the overall sign of the diagram:  it 
has an overall factor of $s$ if it arises from 
an odd permutation
of the outgoing state $\langle {\bf 1'2'...}|$ in 
 the cluster expansion 
of $\langle {\bf 1'2'...}  | T | {\bf 12...} \rangle$. 

\no (vii) Divide by $-\beta$.

\begin{figure}[htb] 
\begin{center}
\hspace{-15mm}
\psfrag{A}{$=2 \pi \deltaE_{\bf 12..m; 12..n} ~ \CM_{\bf 12..m; 12..n}$ }
\psfrag{B}{$=f_0 (\kvec ) /2\om\kvec $}
\psfrag{k}{$\kvec$}
\includegraphics[width=3cm]{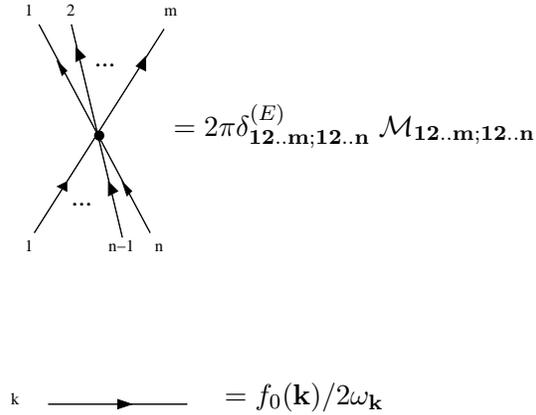} 
\end{center}
\caption{Diagrammatic ingredients} 
\vspace{-2mm}
\label{Figure1} 
\end{figure}

The terms in eq. (\ref{Low.3}) are then represented by the diagram in
Figure 2.   The two terms in eqns. (\ref{Low.5.a}, \ref{Low.5.b}) 
are represented as diagrams in Figure 3.

\begin{figure}[htb] 
\begin{center}
\hspace{-15mm}
\psfrag{A}{$=2 \pi \deltaE_{\bf 12..m; 12..n} ~ \CM_{\bf 12..m; 12..n}$ }
\psfrag{B}{$=\frac{f_0 (\kvec )}{2\om\kvec}$}
\psfrag{k}{$\kvec$}
\includegraphics[width=3cm]{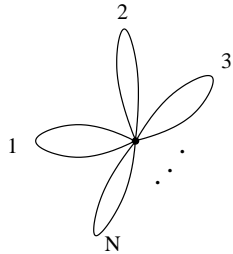} 
\end{center}
\caption{Diagrammatic representation of terms in eq. (\ref{Low.3}) } 
\vspace{-2mm}
\label{Figure2} 
\end{figure}

\begin{figure}[htb] 
\begin{center}
\hspace{-15mm}
\psfrag{A}{$=2 \pi \deltaE_{\bf 12..m; 12..n} ~ \CM_{\bf 12..m; 12..n}$ }
\psfrag{B}{$=\frac{f_0 (\kvec )}{2\om\kvec}$}
\psfrag{k}{$\kvec$}
\includegraphics[width=2cm]{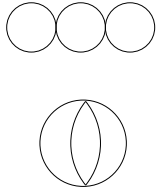} 
\end{center}
\caption{Diagrammatic representation of the two terms in 
eq. (\ref{Low.5.a}) and
(\ref{Low.5.b}) respectively.} 
\vspace{-2mm}
\label{Figure3} 
\end{figure}

The structure of the ``ring'' diagrams shown in Figure 4 
are especially simple.    We compute the sum of such diagrams 
in order to illustrate the above rules and also since we
will need the result in the next section.   Let the diagram
in Figure 4  be denoted as $\CF^{(n)}_{\rm ring}$.   The symmetry
factor of this diagram is $2$ for $n=1$ and $n$ for $n\geq 2$. 
For fermions,  the overall sign of the diagram is $s^{n+1}$.  
The sum over $n$ then simply gives a $\log$:
\barray
\nonumber
\sum_{n=2}^\infty  \CF^{(n)}_{\rm ring} &=& - \inv{\beta} \int \dkind\kvec 
\( -\frac{y_0 (\kvec)}{2} - s\log ( 1 - sy_0 (\kvec ) ) \) 
\\
\label{ringdiagrams}
y_0 (\kvec) \equiv &=&  \frac{\beta f_0 (\kvec)}{2\om\kvec} 
\int \dkind{\kvec'} ~ \CM_{\bf 12;12} (\kvec , \kvec') 
\frac{f_0 (\kvec')}{2\om{\kvec'}}
\earray

\begin{figure}[htb] 
\begin{center}
\hspace{-15mm}
\includegraphics[width=3cm]{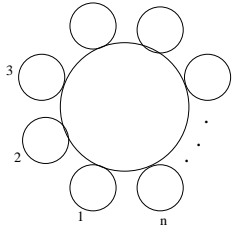} 
\end{center}
\caption{Ring diagrams.} 
\vspace{-2mm}
\label{Figure4} 
\end{figure}

\section{$\Fhat$ and the saddle point  equations}

We now return to including interactions in  the
free energy functional $\Fhat$ of section II.  
Let us write
\beq
\label{var.1}
\Fhat (f,f_0) = \Fhat_0 (f,f_0) + \Fhatone
\eeq
where $\Fhat_0$ is given in eq. (\ref{L10}) and we define
$\U$ as the ``potential'' which incorporates interactions:
\beq
\label{var.2}
\Fhatone = -\inv{\beta} 
\int \dkind\kvec ~ \U (\kvec )
\eeq

It is not too difficult to understand that $\Fhatone$ 
is given by the 2-particle irreducible diagrams of
the last section.  These are defined as diagrams that
cannot become disconnected by the cutting of 2 internal lines. 
For instance,  starting from the $N=2$ diagram in Figure 2,
it is clear that the ring diagrams can generated by attaching
additional loops.  The ring diagrams are not 2-particle irreducible
and should thus not be included in $\Fhatone$.  A more detailed argument was given
by  Lee and Yang\cite{LeeYang}. 
We express this explicitly as follows:   
$\Fhatone $ is just $\CF_{\rm 2-part.~ irred.} (f_0)$ with $f_0$ replaced by $f$:
\beq
\label{fhat1}
\Fhatone (f) = \sum \CF_{\rm 2-part. irred.} (f_0 \to f) 
\eeq
This is shown in Figure 4,  where it is implicit that the lines
have a factor of $f$ rather than $f_0$.  
\begin{figure}[htb] 
\begin{center}
\hspace{-15mm}
\psfrag{F}{$\Fhatone = $}
\includegraphics[width=10cm]{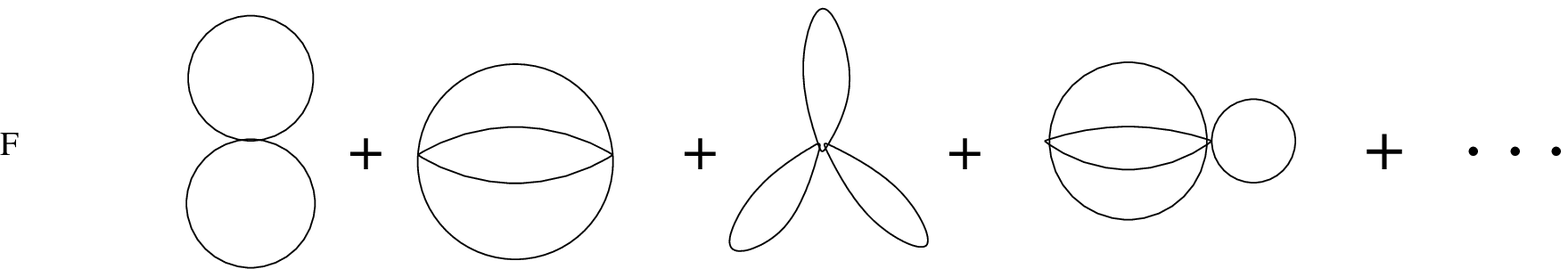} 
\end{center}
\caption{$\Fhat_1$ as the sum of 2 particle irreducible diagrams.} 
\vspace{-2mm}
\label{Figure5} 
\end{figure}

Given $\Fhat$,  $f$ is determined by the saddle point equation:
\beq
\label{var.3}
\log \( \frac{1+sf}{f} \) - \log \( \frac{1+sf_0}{f_0} \) =  -
\frac{\d \U}{\d f} 
\eeq
Substituting the solution of (\ref{var.3}) back into $\Fhat$, it
can be written in a variety of ways, depending on how one utilizes
the relation (\ref{var.3}).  One useful way is:
\barray
\label{var.4}
\CF &=& -\inv{\beta} \int \dkind\kvec ~ 
\[ s \log (1+sf) + ( 1-f\d_f   ) \U \] 
\\ \nonumber
&=& \CF_0 - \inv{\beta} \int \dkind\kvec 
\[ s \log \( \frac{1+sf}{1+ s f_0} \) + ( 1-f\d_f ) \U \]
\earray

It is convenient to define a pseudo-energy $\vep$ as the
following parameterization of $f$:
\beq
\label{pseudo.1}
f \equiv \inv{ e^{\beta \vep} -s }
\eeq
Then the saddle point equation and free energy density take the form:
\barray
\label{pseudo.2}
\vep &=& \omega - \mu - \inv{\beta} \frac{\d \U}{\d f} 
\\ 
\label{pseudo.3}
\CF &=& - \inv{\beta} \int  \dkind{\kvec} 
\[ -s \log (1 - s e^{-\beta \vep} ) 
+ (1- f\d_f  ) \U  \]
\earray

\section{Two-body approximation}

If the gas is not too dense, one expects that 2-particle scattering
will give the most  important contribution.   This is especially
true in the non-relativistic case where scattering preserves the number
of particles.  

Define the following kernel from the 2-particle scattering amplitude:
\beq
\label{2bod.1}
\K (\kvec , \kvec' ) \equiv \inv{4 \om\kvec \om{\kvec'} } 
\CM_{\bf 12;12} (\kvec, \kvec' ) 
\eeq
and the convolution:
\beq
\label{2bodstar}
(\K * f )(\kvec) \equiv  \int \dkind{\kvec'} ~ \K (\kvec, \kvec' ) f(\kvec') 
\eeq
The 2-particle contribution to $U$ can then be written as: 
\beq
\label{2bod.2}
\U (\kvec ) = \frac\beta {2}  f(\kvec )  ~  (\K * f)(\kvec )  
\eeq
The saddle point equation and free energy then take the forms: 
\beq
\label{2bod.3}
\log \( \frac{1+sf}{f} \) - \log \( \frac{1+sf_0}{f_0} \) = - \beta 
~ \K * f 
\eeq
\beq
\label{2bod.4}
\CF = -\inv{\beta} \int \dkind{\kvec} 
\[ s \log (1+sf) + \frac{f}{2}  \log \( \frac{1+sf}{f} \frac{f_0}{1+sf_0} \) 
\]
\eeq

In terms of the pseudo-energy:
\barray
\label{2bod.5}
\vep &=& \omega - \mu -  \K * \( \inv{e^{\beta \vep} -s } \)
\\
\label{2bod.6}
\CF &=&
-\inv{\beta} \int \dkind\kvec 
\[ -s \log (1-s e^{-\beta \vep}) + \frac{\beta f}{2} (\vep - \omega + \mu) \]
\earray

In this two-body approximation,  the integral equation (\ref{2bod.5}) 
resums all diagrams involving two-body scattering.  These ``foam  diagrams''  
are of the kind
shown in Figure 6.

\begin{figure}[htb] 
\begin{center}
\hspace{-15mm}
\psfrag{A}{$=2 \pi \deltaE_{\bf 12..m; 12..n} ~ \CM_{\bf 12..m; 12..n}$ }
\psfrag{B}{$=\frac{f_0 (\kvec )}{2\om\kvec}$}
\psfrag{k}{$\kvec$}
\includegraphics[width=6cm]{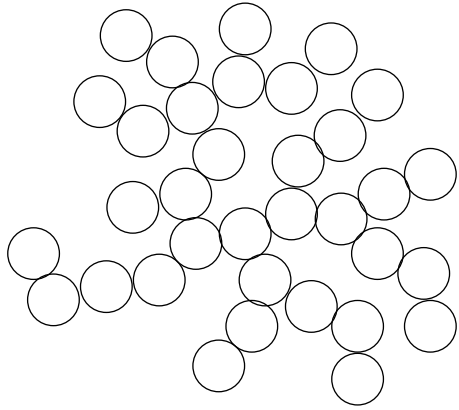} 
\end{center}
\caption{Foam  diagrams.} 
\vspace{-2mm}
\label{Figure6} 
\end{figure}

The foam  diagrams contain much more than the ring diagrams.  As a check, 
we now show how the ring diagrams are contained in the solution of
the integral equation.  Let us define $y$ as 
$\log( \frac{1+sf}{1+sf_0}) = \log (1+y)$ where
\beq
\label{y.1}
y \equiv \frac{ f- f_0}{1+s f_0} 
\eeq
The integral equation (\ref{2bod.3}) can then be expanded in powers of $y$:
\beq
\label{y.2}
y - \frac{(1 + 2s f_0)}{2 f_0} ~  y^2 + .... = \beta \K *(f_0 + (1+sf_0) y ) 
\eeq
Now let $y= y_0 + y_1 + ...$ and plug this into the above equation.  
Many kinds of terms are generated, but to compare with the ring 
diagrams we only focus on terms of the type in eq. (\ref{ringdiagrams}).  
To lowest orders one finds 
\beq
\label{y.3}
y = y_0 + 2 s\,  y_0^2 + ....
\eeq
where $y_0$ is defined in eq. (\ref{ringdiagrams}).   Plugging this lowest
order solution into eq. (\ref{2bod.4}) for the free energy one finds 
\beq
\label{y.4}
\CF - \CF_0 = -\inv{\beta} 
\int  \frac{ d^d \kvec}{ (2\pi)^d } ~ ( y_0 + s\, y_0^2 /2 + ....)
\eeq
which agrees with the low order expansion of eq. (\ref{ringdiagrams}).

\def\rvec{\vec{r}} 
\def\coupling{\gamma}

\section{Non-relativistic case:  hard-core bosons}

We have not assumed the underlying theory is Lorentz invariant 
in the above construction.   However, in the definition eq. (\ref{FS4}) of
the states $|\kvec_1 \cdots \kvec_n \rangle$ we included 
factors of $\sqrt{2 \om\kvec}$ in order that 
$\dklineind\kvec$ in eq. (\ref{2.4})  is Lorentz invariant. 
This simplifies  the comparision with other approaches 
for relativistic models, as we did in section IVC for
the $\phi^4$ theory, since then $\CM$ are the conventional
scattering amplitudes.   

\subsection{Generalities of the non-relativistic case}

For non-relativistic theories, where 
$\om\kvec = \kvec^2 /2m$, 
it is more conventional to normalize
the states as
\beq
\label{nonrel.1}
| \kvec_1 \cdots \kvec_N \rangle = a^\dagger_{\kvec_1} \cdots 
a^\dagger_{\kvec_N} |0\rangle 
\eeq
The formulas of the previous sections still apply but
with the modification:
\beq
\label{nonrel.2}
\dklineind\kvec = \int \frac{d^d \kvec}{(2\pi)^d 2 \om\kvec} 
\to  \int \frac{d^d \kvec}{(2\pi)^d  }
\eeq

We will keep the definition eq (\ref{Type1.1}) of $\CM$ in terms of $\teehat$. 
The formulas for the two-body approximation in section VII then
all apply but now with:
\beq
\label{nonrel.3}
\K (\kvec , \kvec' ) \equiv \CM_{\bf 12,12} (\kvec , \kvec' ) 
\eeq
The above kernel has dimensions of ${\rm energy\cdot volume}$ 
which corresponds to  an energy dimension of $1-d$ (up to velocity
factors). 

\subsection{Hard core boson model}

\def\rvec{{\bf x}}

Consider the two-body potential in position space $\rvec$:
\beq
\label{nonrel.4}
V(\rvec , \rvec' ) = \frac{\coupling}{2} \delta^{(d)} (\rvec - \rvec' )
\eeq
This leads to the second quantized hamiltonian:
\beq
\label{nonrel.5}
H = \int  d^d \rvec 
\( \inv{2m} \vec{\nabla} \psi^\dagger \cdot \vec{\nabla} \psi + 
\frac{\coupling}{4} \psi^\dagger
\psi^\dagger \psi \psi  \)
\eeq

To lowest order $T=H_1$ and 
\beq
\label{nonrel.6}
\langle \alpha' | H_1 | \alpha \rangle = - (2\pi)^d \delta^{(d)} (\kvec_\alpha - \kvec_{\alpha'})
~ \CM_{\alpha' ; \alpha}
\eeq
Expanding the field in terms of annihilation operators:
\beq
\label{nonrel.7}
\psi (\rvec ) = \int \dkind\kvec ~ e^{i\kvec \cdot \rvec} ~ a_\kvec 
\eeq
then to lowest order one finds
\beq
\label{nonrel.8}
M_{\bf 12; 12} = -\coupling  = \kernel
\eeq

The coupling constant $\coupling$ has units of 
${\rm energy} \times {\rm volume}$.  It can be expressed in 
terms of a physical scattering length $a$ as follows. 
To first order in perturbation theory the differential 
cross-section in the center of mass is:
\beq
\label{nonrel.9}
\frac{ d\sigma}{d\Omega} = \frac{m^2 \coupling^2}{4 (2\pi)^{d-1} } ~k^{d-3} 
\eeq
where $k$ is the magnitude of $\kvec$ for one of the incoming particles.  
Since a cross section has dimensions of ${\rm length}^{d-1}$,  we define
$a$ such that the cross section is $a^{d-1}$ when the wavelength of the
particle is $2\pi/a$: 
\beq
\label{nonrel.10}
\frac{ d\sigma}{d\Omega} \Bigr\vert_{k \sim 2\pi /a} \sim a^{d-1} 
\eeq
This leads us to make  the definition:
\beq
\label{nonrel.11}
\frac{\coupling}{(2\pi)^{d/2}} \equiv  \frac{a^{d-2}}{m}
\eeq

\def\hT{h}
\def\zdelta{{z_\delta}}
\def\Li{{\rm Li}}
\def\lambdaT{\lambda_T}

\subsection{Lowest order solution}

It is interesting to carry out our analysis for arbitrary 
spacial dimension $d>0$.  
Integrals over momenta can  be traded for integrals over $\omega$:
\beq
\label{nonrel.12}
\int  \dkind\kvec  =  \( \frac{m}{2\pi} \)^{d/2} \inv{\Gamma(d/2)}  \int_0^\infty d\omega 
~ \omega^{(d-2)/2}
\eeq
For a constant kernel $\kernel = -\coupling$, and $d>0$,
  the solution to the integral 
equation eq. (\ref{2bod.5}) takes the simple form: 
\beq
\label{nonrel.13}
\vep (\kvec) = \om\kvec - \mu + T \delta
\eeq
where
$\delta$ is independent of $\kvec$ and satisfies the equation\footnote{We have used
$\int_0^\infty dx  ~ x^{\nu-1}/ (e^x/z -1 ) = \Gamma (\nu) \Li_\nu (z) ~~ $ for $\Re (\nu) >0$.}
\beq
\label{nonrel.14}
\delta =  \hT  ~  \Li_{d/2} ( z_\mu \zdelta )
\eeq
where we have defined  the fugacities 
\beq
\label{nonrel.15}
 z_\mu  \equiv e^{\beta \mu} , ~~~~~ \zdelta = e^{-\delta} 
\eeq
and a renormalized thermal coupling $\hT$ and thermal wavelength $\lambdaT$:
\beq
\label{nonrel.16}
\hT  \equiv \( \frac{a}{\lambdaT} \)^{d-2},  ~~~~~ \lambdaT  \equiv \inv{ \sqrt{mT} }
\eeq
The function $\Li_\nu (z)$ is the standard polylogarithm, defined as the appropriate
analytic continuation of
\beq
\label{nonrel.17}
\Li_\nu (z) = \sum_{n=1}^\infty \frac{z^n}{n^\nu}
\eeq
The equation (\ref{nonrel.14}) is thus seen as  non-trivial, transcendental  equation 
that determines $\delta$ as a function of $\mu, T$, and the coupling $\hT$. 
Given the solution $\delta (\mu )$ of this equation, using 
eq. (\ref{2bod.6}) the density  can be expressed as
\beq
\label{nonrel.18}
n(\mu )  = \inv{ (2\pi \lambdaT^2 )^{d/2} } \Li_{d/2} ( z_\mu z_{\delta} ) = 
\frac{T \delta }{\coupling}
\eeq
For $d>0$ one can integrate by parts and obtain the following expressions
for the free energy 
\beq
\label{nonrel.18b}
\CF = - \frac{T}{ (2\pi \lambdaT^2 )^{d/2} } 
\( \Li_{(d+2)/2} ( z_\mu z_{\delta} )  + 
\frac{\delta}{2} \Li_{d/2} ( z_\mu z_{\delta}  ) \) 
\eeq

\def\nphys{n_{\rm phys.}}
\def\dtwo{{\textstyle \frac{d}{2} }}
\def\dplus{{ \textstyle \frac{d+2}{2} }}

The case of $2d$ is interesting since the eq. (\ref{nonrel.14}) then becomes
algebraic due to  $\Li_1 (z) = - \log (1-z)$.  
These formulas will be applied to Bose-Einstein condensation, 
and other problems,  in
\cite{ALBose}.

\section{Concluding remarks}

In this paper we have mainly focused on developing the formalism
in a general way for both relativistic and non-relativistic theories.
In the two-body approximation,  the main result is summarized in
the two formulas eqs. (\ref{2bod.5},\ref{2bod.6}) and is quite
straightforward to implement:  one computes, or measures,  the 2-body 
forward-scattering
kernel ${\bf K}$ at zero temperature,  solves the integral equation 
(\ref{2bod.5}), and the free energy is given in terms of this solution
in eq. (\ref{2bod.6}).   For non-constant kernels, one will probably
need to solve the equations numerically.

A formalism that efficiently disentangles zero temperature 
dynamics and quantum statistical sums potentially has many applications,
and we discuss some  of them in these concluding remarks. 

Our formalism is especially well-suited for studies of the effects
of interactions in Bose-Einstein condensation, since there the
filling fractions play a central role.  We have already obtained
some results on this problem and will publish them elsewhere\cite{ALBose}. 
The results presented there also give new insights on 
the Riemann hypothesis.

For high-energy particle physics, our formalism has the following
potential advantages over the Matsubara approach.  The necessity
to renormalize ultra-violet divergences is essentially a zero
temperature problem,  however the Matsubara approach again entangles
the zero-temperature renormalization with the quantum statistical mechanics
and the issue of scheme-dependent results can sometimes be a problem.
In our formalism,  any need for renormalization is carried out at
zero temperature and the S-matrix is expressed in terms of physical
quantities at zero temperature, such as particle masses, etc.  
Our integral equation could also shed some light on the infra-red
problems that are common in the Matsubara approach and also
require special resummations\cite{kapusta,lebellac,Parwani,Braaten,Foam,Blaizot}.  
We emphasize that since the diagrams in this paper have no relation
to finite temperature Feynman diagrams,  our type of resummation is
entirely different than the resummations carried out in these other works. 
Some aspects of the relativistic case will be published 
in \cite{ALrelativistic}.   Our formalism may help to study 
the known results on the free energy of strongly coupled 
supersymmetric gauge theories\cite{Klebanov}.

Our formalism may also be well suited to studies of the quark-gluon
plasma\cite{Rajagopal} currently being studied at RHIC,  since
many of the hadronic cross-sections are known.

\section{Acknowledgments}

I would like to thank G. Mussardo and H. Thacker for discussions and for
drawing my attention to the work \cite{Ma}.

\section{Appendix A}

In this appendix we 
complete the steps leading to the result (\ref{S.1}) 
obtained in \cite{Ma}. 
Starting from eq. (\ref{R8}), 
the next step is to 
try and use the relation $G_0 - G_0^\dagger = - 2\pi i ~ \delta(E-H_0)$ 
so that one can integrate over $E$.   
One  way to do this uses the operators 
$\Omega$:  
\barray
\nonumber
\Omega (E) &\equiv&  G G_0^{-1} = 1 + G_0 T 
\\
\Omega^{-1} (E) &=& G_0 G^{-1} = 1 - G_0 H_1 
\label{R10}
\earray
As we show below, $\Omega$ is the operator that relates free-particle states
to the  in-states of scattering theory.   
Introduce the notation $X \dLR Y \equiv X (\d Y) - (\d X)  Y$.   Then, using
$\d_E G_0 = - G_0^2 $ and $H_1 \Omega = T$ one can
readily show that 
\beq
\label{R11}
\Omega^{-1} \dLR_E  \Omega  = -2 G_0^2 T
\eeq

In order to deal with the complex conjugation needed in eq. (\ref{R8}), 
for any operator $X(E)$ define $X^* (E)$ as simply $X(E)$ but with $i\vep$ replaced
by $-i\vep$.  
This definition gives   $G^*(E) = G(E)^\dagger$,  
$\Omega^* (E) = 1 + G^\dagger H_1 $, and $(\Omega^{-1})^* = 1- G_0^\dagger H_1 $.  
One then has
\beq
\label{R12}
(\Omega^{-1})^* \dLR _E \Omega^* = -2 (G_0^\dagger)^2 T^\dagger
\eeq

The last step is to define
\beq
\label{R13}
S(E) = (\Omega^{-1})^* \Omega (E)
\eeq
Using eq. (\ref{R8}, \ref{R11}), and the cyclicity of the trace,  one obtains
\beq
\label{Amain}
Z = Z_0 + \inv{4 \pi i} \int_0^\infty dE ~ e^{-\beta E} ~ 
\Tr \( S^{-1} \dLR_E S \) 
\eeq

We now demonstrate that the so-called on-shell matrix elements of the
S-operator are the usual S-matrix.  Consider first the operator $\Omega$. 
Inserting a complete set of states $|\alpha'\rangle$ one finds:
\beq
\label{R15}
\Omega (E_\alpha) | \alpha \rangle = | \alpha \rangle + \sum_{\alpha'} 
\inv{E_\alpha - E_{\alpha'} + i \vep} T_{\alpha' \alpha} ~ | \alpha' \rangle 
\eeq
where 
\beq
\label{R16}
T_{\alpha' \alpha} = \langle \alpha' | T(E) |\alpha \rangle , ~~~~~~ {\rm iff~~} E = E_\alpha 
\eeq
The condition $E= E_\alpha$ is the on-shell condition.  
Eq. (\ref{R15}) are the Lippmann-Schwinger equations: 
\beq
\label{Lipp}
|\alpha \rangle_{\rm in} =  \Omega (E_\alpha) | \alpha \rangle 
\eeq
which arise from 
attempting to solve $H |\alpha \rangle_{\rm in} = E_{\alpha} | \alpha \rangle_{\rm in}$
for the in-states (which also have energy $E_\alpha$).  

Turning now to the S-operator,  using identities given above, one can easily establish
\barray
S(E) &=&  1 - 2 \pi i \, \delta (E- H_0) ~ T(E)
\nonumber 
\\
S^{-1} (E) &=& 1 + 2 \pi i \, \delta (E- H_0) ~ T^\dagger (E) 
\label{R17}
\earray
The latter implies
\beq
\label{Son2}
\langle \alpha ' | S(E) | \alpha \rangle = \langle \alpha' | \alpha \rangle 
- 2 \pi i \, \delta(E_\alpha - E_{\alpha'} ) ~ T_{\alpha'  \alpha} 
, ~~~~~~{\rm iff~~}  E = E_\alpha
\eeq
The equation (\ref{Amain}) together with the identification (\ref{R17}) 
of the S-matrix is the main result obtained in \cite{Ma}.


\begin{thebibliography}{99}







\bibitem{kapusta} J. I. Kapusta, 
{\it Finite Temperature Field Theory}, 
Cambridge, 1989

\bibitem{lebellac}  M. Le Bellac, 
{\it Thermal Field Theory},  Cambridge, 1996.

\bibitem{jackiw}  L. Dolan and R. Jackiw,  
Phys. Rev. {\bf D9} (1974)  3320.  

\bibitem{yangyang} C. N. Yang and C. P. Yang, Jour. Math. Phys. {\bf 10},
(1969) 1115. 


\bibitem{ZamoTBA} Al. Zamolodchikov, Nucl. Phys. {\bf B342} (1990) 695. 


\bibitem{Ma}
R. Dashen, S.-K. Ma and H. J. Bernstein,  Phys. Rev. 
{\bf 187} (1969) 345.

\bibitem{Thacker}  H. B. Thacker, 
Phys. Rev. {\bf D16} (1977) 2515.

\bibitem{Ma2}  R. Dashen and S.-K. Ma, 
J. Math. Phys. {\bf 12} (1971) 689. 


\bibitem{LeeYang}
T. D. Lee and C. N. Yang, 
Phys. Rev. {\bf 113} (1959) 1165;    Phys. Rev. {\bf 117} (1960) 22. 

\bibitem{Entropygas}  L. D. Landau and E. M. Lifshitz, 
{\it Statistical Physics}, Pergamon Press (1980). 

\bibitem{Gottfried}
K. Gottfried and T.-M. Yan, 
{\it Quantum Mechanics:  Fundamentals}, 
Spring-Verlag, 2003.

\bibitem{Weinberg}
S. Weinberg,
{\it The Quantum Theory of Fields I}, 
Cambridge University Press 1995.

\bibitem{Peskin} M. E. Peskin and D. V. Schroeder, 
{\it An Introduction to Quantum Field Theory}, 
Addison-Wesley 1995. 



\bibitem{LecSal}
 A. LeClair, F. Lesage, S. Sachdev, and H. Saleur, 
Nucl.Phys. {\bf B482} (1996) 579 cond-mat/9606104.

\bibitem{LecMuss} A. LeClair and G. Mussardo,
Nucl.Phys.{\bf B552} (1999) 624. 



\bibitem{Bugrij}
A. I. Bugrij and V. N. Shadura,
arXiv:hep-th/9510232. 


\bibitem{Parwani}   R. R. Parwani, 
Phys.Rev. {\bf D45}  (1992) 4695; Erratum-ibid. {\bf D48} (1993) 5965,
hep-ph/9204216. 

\bibitem{Braaten} E. Braaten and R. D. Pisarski,
Nucl. Phys. {\bf B337} (1990)  569; Nucl. Phys. {\bf B339} (1990) 310;
 J. Frenkel and J. C. Taylor, 
Nucl. Phys. {\bf B334} (1990) 199. 

\bibitem{Foam}  I. T. Drummond, R. R. Horgan, and P. V. Landshoff, 
Nucl.Phys. {\bf B524}  (1998) 579,
hep-ph/9708426

\bibitem{Blaizot}  J.-P. Blaizot, E. Iancu, and A. Rebhan, 
Phys.Rev. {\bf D63}  (2001) 065003,
hep-ph/0005003. 















\bibitem{ALBose}  A. LeClair, {\it Interacting Bose and Fermi 
gases in low dimensions and the Riemann hypothesis}, 
to appear.  

\bibitem{ALrelativistic}  A. LeClair, {\it Pressure of relativistic
quantum gases from the S-matrix,},  to appear. 




\bibitem{Rajagopal}  M. G. Alford,  K. Rajagopal, and F. Wilczek,
Phys. Lett. {\bf B422} (1998) 247.  


\bibitem{Klebanov}  S. S. Gubser, I. R. Klebanov, and A. A. Tseytlin,
Nucl. Phys. {\bf B534} (1988) 202, hep-th/9805156. 


\end{thebibliography}
\end{document}